\documentclass[11pt,twocolumn,twoside]{article}

\usepackage{./fully3d/fully3d}
\usepackage{amsmath,amsfonts,amssymb,mathrsfs}
\usepackage{bm}
\usepackage{microtype}
\usepackage{mathtools}

\usepackage{algorithm}
\usepackage{algpseudocode}

\usepackage{acro} % acronyms
\acsetup{single}

\usepackage{caption}
\usepackage{subcaption}
\usepackage[labelformat=simple,font=footnotesize]{subcaption}

\usepackage[font=small]{caption}

\usepackage{tikz}
\usetikzlibrary{arrows.meta,shapes,intersections,calc,angles,decorations.pathreplacing,intersections,quotes,angles,positioning,fit,petri,through,shadows,plotmarks,spy}

\usepackage{pgfplots}
\usepackage{pgfplotstable}
\usetikzlibrary{pgfplots.groupplots}

\usepackage{overpic}
\usepackage{color}

\usepackage{float}
\usepackage{flushend} % For column balancing

\usepackage{comment}

\newcommand{\boldr}{\bm{r}}
\newcommand{\bolds}{\bm{s}}

\newcommand{\boldx}{\bm{x}}
\newcommand{\boldy}{\bm{y}}
\newcommand{\boldz}{\bm{z}}

\newcommand{\boldlambda}{\bm{\lambda}}
\newcommand{\boldmu}{\bm{\mu}}
\newcommand{\boldtheta}{\bm{\theta}}
\newcommand{\boldphi}{\bm{\phi}}
\newcommand{\boldpsi}{\bm{\psi}}
\newcommand{\boldzero}{\bm{0}}
\newcommand{\boldepsilon}{\bm{\epsilon}}

\newcommand{\boldI}{\bm{I}}

\newcommand{\boldR}{\bm{R}}

\newcommand{\calN}{\mathcal{N}}

\newcommand{\calX}{\mathcal{X}}
\newcommand{\calY}{\mathcal{Y}}
\newcommand{\calZ}{\mathcal{Z}}

\newcommand{\rme}{\mathrm{e}}

\newcommand{\rml}{\mathrm{l}}

\newcommand{\rmt}{\mathrm{t}}

\newcommand{\ybar}{\bar{y}}

\newcommand{\R}{\mathbb{R}}

\newcommand{\transp}{^\top}
\DeclareAcronym{2D}{
	short=2-D,
	long=2-dimensional,
}

\DeclareAcronym{3D}{
	short=3-D,
	long=3-dimensional,
}

\DeclareAcronym{dD}{
	short=$d$-D,
	long=$d$-dimensional,
}

\DeclareAcronym{mD}{
	short=$m$-D,
	long=$m$-dimensional,
}

\DeclareAcronym{CT}{
	short=CT, 
	long=computed tomography,
}

\DeclareAcronym{PET}{
	short=PET, 
	long=positron emission tomography,
}

\DeclareAcronym{SPECT}{
	short=SPECT, 
	long=single-photon emission CT,
}

\DeclareAcronym{MRI}{
	short=MRI, 
	long=magnetic resonance imaging,
}

\DeclareAcronym{MR}{
	short=MR, 
	long=magnetic resonance,
}

\DeclareAcronym{PCCT}{
	short=PCCT, 
	long=photon-counting computed tomography,
}

\DeclareAcronym{DECT}{
	short=DECT, 
	long=dual-energy computed tomography,
}

\DeclareAcronym{MBIR}{
	short=MBIR, 
	long=model-based iterative reconstruction,
}

\DeclareAcronym{WLS}{
	short=WLS, 
	long=weighted least squares,
}
\DeclareAcronym{PWLS}{
	short=PWLS, 
	long=penalized weighted least squares,
}

\DeclareAcronym{PLS}{
	short=PLS, 
	long=parallel level set,
}

\DeclareAcronym{ML}{
	short=ML, 
	long=maximum likelihood,
}

\DeclareAcronym{PML}{
	short=PML, 
	long=penalized maximum likelihood,
}

\DeclareAcronym{MLAA}{
	short=MLAA, 
	long=maximum likelihood activity and attenuation,
}

\DeclareAcronym{SPS}{
	short=SPS, 
	long=separable paraboloidal surrogates,
}

\DeclareAcronym{CS}{
	short=CS,
	long=compressive sensing,
}

\DeclareAcronym{TV}{
	short=TV,
	long=total variation,
}

\DeclareAcronym{TNV}{
	short=TNV, 
	long=total nuclear variation,
}

\DeclareAcronym{JTV}{
	short=JTV, 
	long=joint total variation,
}

\DeclareAcronym{DTV}{
	short=DTV, 
	long=directional total variation,
}

\DeclareAcronym{SQS}{
	short=SQS, 
	long=separable quadratic surrogate,
}

\DeclareAcronym{ADMM}{
	short=ADMM, 
	long=alternating direction method of multipliers,
}

\DeclareAcronym{DL}{
	short=DL,
	long= deep learning,
}

\DeclareAcronym{DiL}{
	short=DiL,
	long= dictionary learning,
}

\DeclareAcronym{CDL}{
	short=CDL,
	long=convolutional DiL,
}

\DeclareAcronym{MCDL}{
	short=CDL,
	long=multichannel convolutional dictionary learning,
}

\DeclareAcronym{CAOL}{
	short=CAOL, 
	long=convolutional analysis operator learning,
}

\DeclareAcronym{MCAOL}{
	short=MCAOL, 
	long=multichannel convolutional analysis operator learning,
}

\DeclareAcronym{PDF}{
	short=PDF,
	long=probability distribution function,
}

\DeclareAcronym{PSNR}{
	short=PSNR, 
	long=peak signal-to-noise ratio,
}

\DeclareAcronym{SSIM}{
	short=SSIM, 
	long=structural similarity index measure,
}

\DeclareAcronym{SNR}{
	short=SNR, 
	long=signal-to-noise ratio,
}

\DeclareAcronym{CNN}{
	short=CNN, 
	long=convolutional NN,
}

\DeclareAcronym{NN}{
	short=NN, 
	long=neural network,
}

\DeclareAcronym{GAN}{
	short=GAN, 
	long=generative adversarial network,
}

\DeclareAcronym{WGAN}{
	short=W-GAN, 
	long=Wasserstein GAN,
}

\DeclareAcronym{VAE}{
	short=VAE, 
	long=variational autoencoder,
}

\DeclareAcronym{MVAE}{
	short=MVAE, 
	long=multi-branch VAE,
}

\DeclareAcronym{beta-VAE}{
	short=$\beta$-VAE, 
	long=beta-variational autoencoder,
}

\DeclareAcronym{LOR}{
	short=LOR,
	long=line of response,
	long-plural-form = lines of response,
}

\DeclareAcronym{TOF}{
	short=TOF,
	long=time-of-flight,
}

\DeclareAcronym{MAP}{
	short=MAP,
	long=maximum \emph{a~posteriori},
}

\DeclareAcronym{EM}{
	short=EM,
	long=expectation-maximization,
}

\DeclareAcronym{MLEM}{
	short=MLEM,
	long=maximum-likelihood expectation-maximization,
}

\DeclareAcronym{OSEM}{
	short=OSEM, 
	long=ordered subsets expectation maximization,
}

\DeclareAcronym{MAPEM}{
	short=MAPEM, 
	long=maximum \textit{a posteriori} expectation maximization,
}

\DeclareAcronym{FBP}{
	short=FBP, 
	long=filtered backprojection,
}

\DeclareAcronym{IFFT}{
	short=IFFT, 
	long=inverse fast Fourier transform,
}

\DeclareAcronym{GT}{
	short=GT,
	long=ground truth,
}

\DeclareAcronym{HU}{
	short=HU,
	long=Houndsfield Units,
}

\DeclareAcronym{LAC}{
	short=LAC,
	long=linear attenuation coefficient,
}

\DeclareAcronym{AC}{
	short=AC,
	long=attenuation coefficient,
}

\DeclareAcronym{MNIST}{
	short=MNIST,
	long=Modified National Institute of Standards and Technology,
}

\DeclareAcronym{LBFGS}{
	short=L-BFGS,
	long=limited-memory  Broyden-Fletcher-Goldfarb-Shanno,
}

\DeclareAcronym{KL}{
	short=KL,
	long=Kullback-Leibler
}

\DeclareAcronym{ReLU}{
	short=RelU,
	long=rectified linear unit
}

\DeclareAcronym{PSO}{
	short=PSO,
	long=particle swarm optimization
}

\DeclareAcronym{DM}{
	short=DM,
	long=diffusion model
}

\DeclareAcronym{DPS}{
	short=DPS,
	long=diffusion posterior sampling
}

\DeclareAcronym{PCA}{
	short=PCA,
	long=principal component analysis 
}
\DeclareAcronym{MSE}{
	short=MSE,
	long=mean squared error 
}

\DeclareAcronym{XCAT}{
	short=XCAT,
	long=extended cardiac-torso,
}

\DeclareAcronym{OOD}{
	short=OOD,
	long=out-of-distribution,
}

\DeclareAcronym{FWHM}{
	short=FWHM,
	long=full width at half maximum,
}

\DeclareAcronym{PVE}{
	short=PVE,
	long=partial volume effect,
}

\DeclareAcronym{DDPM}{
	short=DDPM,
	long=denoising diffusion probabilistic model,
}

\DeclareAcronym{JRAA}{
	short=JRAA,
	long=joint reconstruction of the activity and the attenuation,
}

\usepackage{fontsize}
\changefontsize{10.45} % automatic baselineskip adjustment when specifying fontsize
\AtBeginDocument{%
    \setlength\intextsep{10pt}%
	\setlength\textfloatsep{10pt}}

\begin{document}

%-------------------------------------------------------------------------------------------
%%%%% add your title here %%%%%
\title{Joint Reconstruction of the Activity and the Attenuation  in PET  by Diffusion Posterior Sampling: a Feasibility Study} 

%%%%% add authors and affiliations here %%%%%
\author[1]{Cl\'ementine Phung-Ngoc}
\author[1]{Alexandre Bousse}
\author[1]{Antoine De Paepe}
\author[3]{Hong-Phuong Dang}
\author[2]{Olivier Saut}
\author[1]{Dimitris Visvikis}

\affil[1]{Univ. Brest, LaTIM, INSERM, UMR 1101, 29238 Brest, France.}
\affil[2]{INRIA Monc, Université de Bordeaux, Bordeaux INP, CNRS, 33405 Talence, France.}
\affil[3]{IETR - UMR CNRS 6164, CentraleSupélec Rennes Campus, 35576 Cesson-Sevigné, France.}
%%%%% don't change these 2 lines %%%%%
\maketitle
%\thispagestyle{fancy}

%-------------------------------------------------------------------------------------------
%%%%% add your summary (abstract) here               %%%%%%
%%%%% use footnotesize for this section              %%%%%%
%%%%% please stick to the customabstract environment %%%%%% 

\begin{customabstract}
	This study introduces a novel framework for \ac{JRAA} in \ac{PET} using \ac{DPS}. By leveraging \acp{DM}, this approach directly addresses activity--attenuation dependencies, mitigating crosstalk issues prevalent in non-\ac{TOF} settings. Experimental evaluations, conducted using \ac{2D} \acs{XCAT} phantom data, demonstrate that \ac{DPS} significantly outperforms traditional \ac{MLAA} methods, producing consistent and high-quality reconstructions even in the absence of \ac{TOF} information. Ongoing work aims to extend our method to real \ac{3D} data with encouraging preliminary findings.
\end{customabstract}

\acresetall

%-------------------------------------------------------------------------------------------
%%%%% main text                                            

\section{Introduction}

\Ac{PET} plays a critical role in medical imaging such as in oncology and cardiology. For accurate quantification, \ac{PET} imaging requires attenuation correction, typically achieved using complementary modalities like \ac{CT} or \ac{MRI}. %However, these additional scans can introduce higher radiation exposure or increase the complexity of the imaging workflow.
However, these additional scans increase radiation exposure and the complexity of the imaging workflow.

To mitigate radiation dose---especially in scenarios such as follow-up studies or potential screening protocols---it is desirable to perform \ac{PET} acquisition in a low-dose setting. A promising approach to achieve this is to derive the attenuation correction directly from the \ac{PET} emission data with \ac{JRAA} techniques. Methods like \ac{MLAA} have demonstrated reasonable success in \ac{TOF} \ac{PET} settings by simultaneously estimating activity and attenuation using  \cite{rezaei2012simultaneous} or by joint estimation of the activity and \acp{AC} \cite{defrise2012time,rezaei2014ml,berker2016attenuation}. However, these methods struggle in non-\ac{TOF} \ac{PET} scenarios due to activity--attenuation crosstalk.

Recent advances in \ac{DL} have opened new possibilities for enhancing \ac{PET} reconstruction. Techniques leveraging \ac{DL} can predict attenuation-corrected images from non-corrected data or directly generate attenuation maps \cite{chen2023deep}. 

In parallel, \acp{DM} have emerged as a powerful tool for solving inverse problems through \ac{DPS} \cite{chung2023diffusion}. These approaches have demonstrated promising performance in medical imaging tasks \cite{webber2024diffusion} and more particularly in  \ac{PET} reconstruction \cite{singh2023score,webber2024likelihood}.

In this paper, we propose a novel framework for \ac{JRAA} directly from \ac{PET} emission data using \ac{DPS}. Our approach extends the \ac{MLAA} algorithm by integrating the joint prior \ac{PDF} of the activity and attenuation maps derived through \ac{DPS}.

%\textbf{}The rest of the paper is organized as follows: 
Section~\ref{sec:method} introduces the forward model, the inverse problem, the \ac{DPS} approach from %\citeauthor{chung2023diffusion}
~\cite{chung2023diffusion} and its integration into an \ac{MLAA}-like framework. Section~\ref{sec:results} presents experimental results using the \acs{XCAT} phantom \cite{segars20104d} for both \ac{TOF} and non-\ac{TOF} data. Section~\ref{sec:discussion} discusses the limitations of our method and potential avenues for future research. Section~\ref{sec:conclusion} concludes the paper.

\section{Materials and Methods}\label{sec:method}

\subsection{Background on Joint Activity and Attenuation Reconstruction}

%\textcolor{red}{todo: change $\boldz$ to $\boldepsilon$} 

The task consists of retrieving an activity image (or volume) $\boldlambda ~=~[\lambda_1,\dots,\lambda_m]\transp \in \calX \triangleq \R_+^m$, $m$ being the number of pixels (or voxels), from a collection of \ac{PET} measurements $\boldy = \{y_{i,k}\} \in \calY \triangleq \R^{n_{\rml}\cdot n_{\rmt}}$ where  $y_{i,k}$ denotes the number of detected $\gamma$-photon pairs at the $i$th \ac{LOR} and $k$th time bin (for \ac{TOF} \ac{PET}), $n_{\rml}$ and  $n_{\rmt}$ being respectively the number of \acp{LOR} and the number of time bins ($n_{\rmt}=1$ in the non-\ac{TOF} case). %The reconstruction of $\boldlambda$ from $\boldy$ 
The counting process is modeled with a Poisson random \ac{PDF},
\begin{equation}\label{eq:poisson}
	y_{i,k} \mid \boldlambda,\boldmu \sim \mathrm{Poisson} ( \ybar_{i,k} (\boldlambda,\boldmu) )
\end{equation} 
where the expectation $\ybar_{i,k} (\boldlambda,\boldmu)$ depends on the activity image $\boldlambda$ to reconstruct as well as on the $\gamma$-photon attenuation map represented by an image $\boldmu = [\mu_1,\dots,\mu_m]\transp\in \calX$. Ignoring background effects such as scatter and random coincidences, a standard forward model is
\begin{equation}\label{eq:forward}
	\ybar_{i,k}(\boldlambda,\boldmu) \triangleq  a_i(\boldmu) \cdot \sum_{j=1}^m p_{i,k,j} \lambda_j  
\end{equation}
where $p_{i,k,j}$ is the probability that an emission from voxel $j$ is detected in $(i,k)$ (taking into account the system's geometry, resolution  and sensitivity) and $a_i(\boldmu)$ is the $i$th \ac{AC}, given by the Beer-Lambert law as 
\begin{equation}\label{eq:attn}
	a_i(\boldmu)  = \rme^{- [\boldR \boldmu]_i } \, ,
\end{equation}
$\boldR \in \R^{n\times m}$ being the discrete Radon transform which computes line integrals along each \ac{LOR}. 

In multimodal imaging, the attenuation map $\boldmu$ can be derived from an anatomical image obtained from \ac{CT} or \ac{MRI}. The activity image $\boldlambda$ can then be reconstructed by solving the \ac{MAP} optimization problem
\begin{equation}\label{eq:map}
	\max_{\boldlambda\in\calX} \, p(\boldy \mid \boldlambda,\boldmu )  \cdot p(\boldlambda)  
\end{equation}
where the conditional \ac{PDF} $p(\boldy | \boldlambda,\boldmu )$ is given by \eqref{eq:poisson} (assuming conditional independence of the measurements) and $p(\boldlambda)$ is the prior distribution on $\boldlambda$. Since $p(\boldlambda)$ is unknown, it is typically approximated by $p(\boldlambda) = \exp(-R(\boldlambda))$ where $R\colon \calX \to \R^+$ is a convex regularizer which ideally promotes image smoothness while preserving edges, in which case \eqref{eq:map} is a \ac{PML} problem that can be solved by means of iterative algorithms \cite{depierro1995}.

When $\boldmu$ is not available, it can be jointly estimated  with $\boldlambda$ from the emission data $\boldy$ by solving     
\begin{equation}\label{eq:map2}
	\max_{\boldx\in\calX^2} \, p(\boldy \mid \boldx )  \cdot p(\boldx)  
\end{equation}
where $\boldx = (\boldlambda,\boldmu)$ is the two-channel image comprising the activity and the attenuation. In absence of the prior $p(\boldx)$, solving \eqref{eq:map2} corresponds to the \ac{MLAA} problem which can be solved using an iterative algorithm \cite{rezaei2012simultaneous}. However, the results are affected by the $\boldlambda$--$\boldmu$ crosstalk when the \ac{TOF} resolution is too low. Moreover, this approach can only reconstruct the images up to a scaling factor due to the $\ybar_{i,k}(\boldlambda/c,\boldmu -\log c) = \ybar_{i,k}(\boldlambda,\boldmu)$ identity.

\subsection{Joint Activity and Attenuation Reconstruction using Diffusion Posterior Sampling}

\Ac{JRAA} can leverage the utilization of the \ac{PDF} $p(\boldx)$ in \eqref{eq:map2} to take into account prior  knowledge on the pair $(\boldlambda,\boldmu)$. This \ac{PDF} being unknown, we propose to sample $\boldx$ from $p(\boldx | \boldy)$ using \ac{DPS} \cite{chung2023diffusion}. In the following $\calZ = \calX \times\calX$ denotes the activity/attenuation map space.

In \acp{DM}, noise is incrementally added to an image $\boldx_0$, sampled from the training dataset with a \ac{PDF} $p^\mathrm{data}$, through a diffusion process, resulting in a noisy image $\boldx_t$ at each time step $t = 1,\dots,T$. The reverse process   samples an image from a generalized version of $p^\mathrm{data}(\boldx)$, approximating the theoretical prior $p(\boldx)$. We used the \ac{DDPM} \cite{ho2020denoising} which  samples   $\boldx_t$ given $\boldx_{t-1}$ as
\begin{equation}\label{eq:xtfromx0}
	%\boldx_t \sim \calN(\sqrt{\bar{\alpha}_t} \boldx_0, (1-\bar{\alpha}_t)\boldI),
	\boldx_t \mid \boldx_{t-1}  \sim \mathcal{N} \left( \sqrt{\alpha_t} \boldx_{t-1} , (1-\alpha_t) \boldI_{\calZ}  \right)
\end{equation}
where $\boldI_{\calZ}$ is the identity matrix on $\calZ$ and $\alpha_t$ is a scaling factor defined such that $\boldx_T \sim \calN (\boldzero_{\calZ},\boldI_{\calZ})$. An approximate reverse process, involving the score function $\bolds(\boldx_t, t) \triangleq \nabla \log p_t(\boldx_t)$---$p_t$ being the \ac{PDF} of $\boldx_t$---can be derived to compute $\boldx_{t-1}$ from $\boldx_t$ as
\begin{align}\label{eq:sampling}
	\boldx_{t-1} & = \frac{\sqrt{\alpha_t} (1-\bar{\alpha}_{t-1})}{1-\bar{\alpha}_t} \boldx_t + \frac{\sqrt{\bar{\alpha}_{t-1}} \beta_t}{1 - \bar{\alpha}_t} \hat{\boldx}_0(\boldx_t) + \sigma_t \boldz \nonumber \\
	\boldz & \sim \calN(\boldzero_{\calZ},\boldI_{\calZ}) \, ,
\end{align}
where $\bar{\alpha}_t = \prod_{s=1}^t \alpha_s$, $\sigma_t= (1-\alpha_t)(1-\bar{\alpha}_{t-1})/(1-\bar{\alpha}_{t})$, $\beta_t = 1-\alpha_t$ and $\hat{\boldx}_0(\boldx_t) \triangleq \mathbb{E}[\boldx_0 | \boldx_t]$ is given by Tweedie's formula:
\begin{equation}
	\hat{\boldx}_0 (\boldx_t) =  \frac{1}{\sqrt{\bar{\alpha}_t}} (\boldx_t + (1 - \bar{\alpha}_t) \bolds (\boldx_t, t) ) \, .
\end{equation}
The score function $\bolds$ is unknown and  therefore is approximated by a \ac{NN} $\bolds_{\boldtheta} \colon \calZ \times [0,T]\to \calZ$ trained by score matching as
\begin{equation}\label{eq:score_matching}
	\min_{\boldtheta} \, \mathbb{E}_{t, \boldx_0, \boldepsilon} \left[ \left\| \bolds_{\boldtheta}(\boldx_t,t) - \nabla_{\boldx_t} \log(p_t(\boldx_t\mid\boldx_0) \right\|_2^2 \right]
\end{equation}
where $\boldx_0 \sim p^\mathrm{data}$, $\boldepsilon \sim \calN(\boldzero_{\calZ},\boldI_{\calZ})$ and  $\boldx_t = \sqrt{\bar{\alpha}_t}\boldx_0 + \sqrt{1-\bar{\alpha}_t} \boldepsilon$. An equivalent formulation for \acp{DDPM} can be to predict the added noise $\boldepsilon$ \cite{ho2020denoising}. In the following, this analogous formulation is used and $\bolds_{\boldtheta}(\boldx_t,t)$ represents the \ac{NN} trained on noise estimation.

\Ac{DPS} aims at solving an inverse problem  by combining the prior \ac{PDF} $p(\boldx)$ and the likelihood of the measurements $p(\boldy | \boldx)$. It uses the conditional score $\nabla_{\boldx_t} p_t(\boldx_t | \boldy)$ to guide the reverse diffusion process in order to recover an image matching the measurements $\boldy$. The conditional score is given by the Bayes' formula as 
\begin{equation}\label{eq:bayes}
    \nabla_{\boldx_t} \log p(\boldx_t \mid \boldy) = \nabla \log p(\boldx_t) + \nabla_{\boldx_t} \log p(\boldy \mid \boldx_t) \, .
\end{equation}
There is no explicit relation between $\boldy$ and $\boldx_t$, thus making the computation of $p(\boldy | \boldx_t)$ intractable. Instead, the following approximation is commonly used:
\begin{equation}\label{eq:laplaceext}
    \nabla_{\boldx_t} \log p(\boldy \mid \boldx_t) \simeq \nabla_{\boldx_t} \log p(\boldy \mid \hat{\boldx}_0(\boldx_t))\, .
\end{equation}

The overall \ac{DPS} process, which consists in iteratively sampling $\boldx'_{t-1}$ from $\boldx_{t}$ using \eqref{eq:sampling} then performing a gradient descent step from $\boldx'_{t-1}$ using \eqref{eq:laplaceext}, is summarized in Algorithm~\ref{algo:dps}.

\begin{algorithm}\label{alg:dps}
	\caption{DPS for \ac{JRAA}}\label{algo:dps}
	\begin{algorithmic}[1]
		\Require $T$, $\boldy$, $\{ \zeta_t \}_{t=1}^T$, $\{ \xi_t \}_{t=1}^T$, $\{ \sigma_t \}_{t=t}^T$, $\{ \alpha_t \}_{t=t}^T$
		\State $\boldx_T = (\boldlambda_T, \boldmu_T) \sim \calN(\boldzero_{\calZ}, \boldI_{\calZ})$
		\For{$t = T$ \textbf{to} $1$}
		    %\State $\hat{\bolds} \gets \bolds_{\theta}(\boldx_t, t)$
		    \State $\hat{\boldx}_0 \gets \frac{1}{\sqrt{\bar{\alpha}_t}} \big(\boldx_t + \sqrt{1 - \bar{\alpha}_t} \bolds_{\theta}(\boldx_t, t)\big)$
		    \State $\boldz \sim \calN(\boldzero_{\calZ}, \boldI_{\calZ})$
		    \State $\boldx'_{t-1} \gets \frac{\sqrt{\alpha_t} (1-\bar{\alpha}_{t-1})}{1-\bar{\alpha}_t} \boldx_t + \frac{\sqrt{\bar{\alpha}_{t-1}} \beta_t}{1 - \bar{\alpha}_t} \hat{\boldx}_0 + \sigma_t \boldz$
		    \State $(\boldlambda'_{t-1}, \boldmu'_{t-1}) \gets \boldx'_{t-1}$
		    \State $\boldlambda_{t-1} \gets \boldlambda'_{t-1} + \zeta_t \nabla_{\boldlambda_t} \log p(\boldy | \hat{\boldx}_0)$
		    \State $\boldmu_{t-1} \gets \boldmu'_{t-1} + \xi_t \nabla_{\boldmu_t} \log p(\boldy | \hat{\boldx}_0)$
		    \State $\boldx_{t-1} \gets (\boldlambda_{t-1}, \boldmu_{t-1})$
		
		\EndFor
		\State \textbf{return} $\boldx_0 = (\boldlambda_0,\boldmu_0)$
	\end{algorithmic}
\end{algorithm}

\newlength{\tempdima}
\setlength{\tempdima}{0.255\linewidth}

\newlength{\tempdimb}
\setlength{\tempdimb}{0.2\linewidth}

\newcommand{\textcolormetrics}{lightgray}

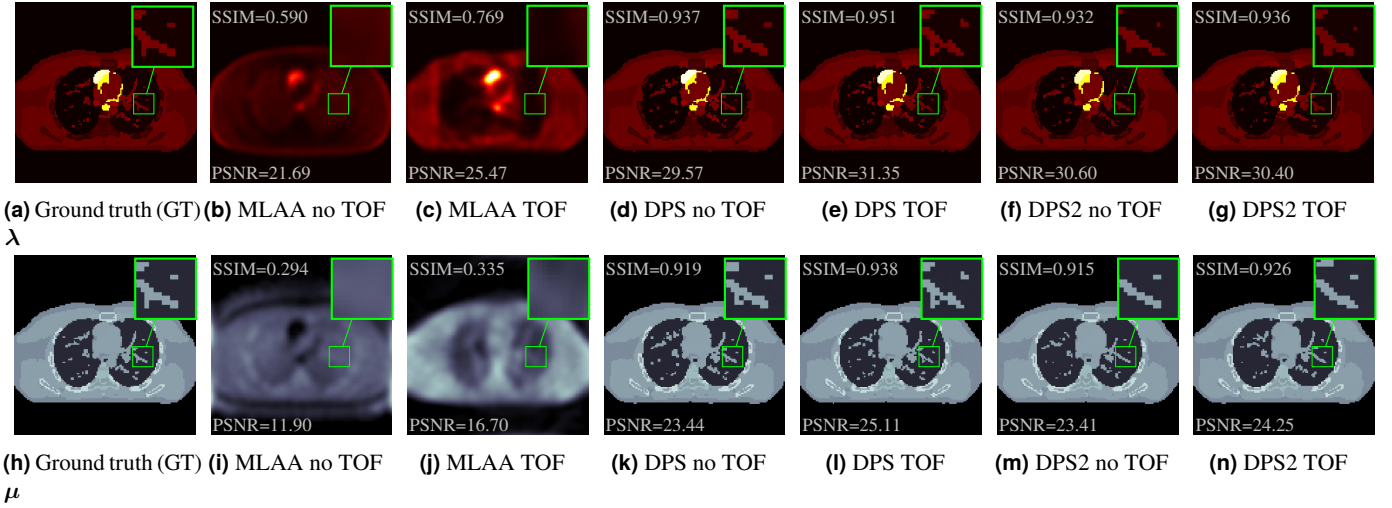
\begin{figure*}[htbp]
	\centering
	
	%%%%% Activity images
	\subfloat[\Ac{GT} $\boldlambda$ \label{subfig:gt_act_hc}]{
		\begin{tikzpicture}
			\begin{scope}[spy using outlines={rectangle,magnification=3,size=8mm,connect spies}]
				\node[inner sep=0pt] {
					\includegraphics[width=\tempdima,height=\tempdima]{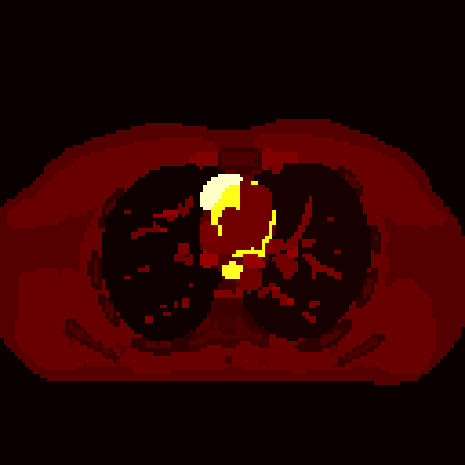}
				};
				\spy [green] on (0.5,-0.15) in node [left,green] at (1.15,0.74);
			\end{scope}
		\end{tikzpicture}
	}
	\subfloat[MLAA no TOF \label{subfig:mlaa_act_notof_hc}]{
		\begin{tikzpicture}
			\begin{scope}[spy using outlines={rectangle,magnification=3,size=8mm,connect spies}]
				\node[inner sep=0pt] {
					\begin{overpic}[width=\tempdima,height=\tempdima]{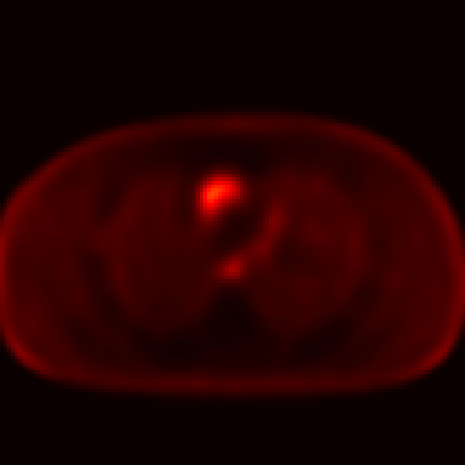}
						\put(1,2){\textcolor{\textcolormetrics}{\scriptsize \acs{PSNR}={21.69}}}
						\put(1,89){\textcolor{\textcolormetrics}{\scriptsize \acs{SSIM}={0.590}}}
					\end{overpic}
				};
				\spy [green] on (0.5,-0.15) in node [left,green] at (1.2,0.75);
			\end{scope}
		\end{tikzpicture}
	}
	\subfloat[MLAA TOF \label{subfig:mlaa_act_tof_hc}]{
		\begin{tikzpicture}
			\begin{scope}[spy using outlines={rectangle,magnification=3,size=8mm,connect spies}]
				\node[inner sep=0pt] {
					\begin{overpic}[width=\tempdima,height=\tempdima]{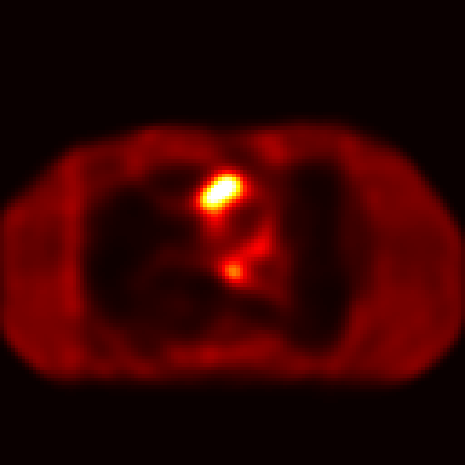}
						\put(1,2){\textcolor{\textcolormetrics}{\scriptsize \acs{PSNR}={25.47}}}
						\put(1,89){\textcolor{\textcolormetrics}{\scriptsize \acs{SSIM}={0.769}}}
					\end{overpic}
				};
				\spy [green] on (0.5,-0.15) in node [left,green] at (1.2,0.75);
			\end{scope}
		\end{tikzpicture}
	}
	\subfloat[DPS no TOF \label{subfig:dps_act_notof_hc}]{
		\begin{tikzpicture}
			\begin{scope}[spy using outlines={rectangle,magnification=3,size=8mm,connect spies}]
				\node[inner sep=0pt] {
					\begin{overpic}[width=\tempdima,height=\tempdima]{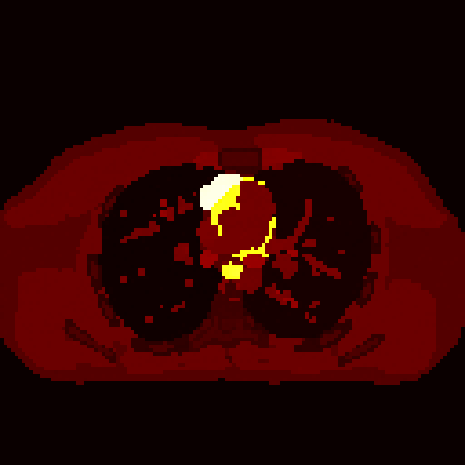}
						\put(1,2){\textcolor{\textcolormetrics}{\scriptsize \acs{PSNR}={29.57}}}
						\put(1,89){\textcolor{\textcolormetrics}{\scriptsize \acs{SSIM}={0.937}}}
					\end{overpic}
				};
				\spy [green] on (0.5,-0.15) in node [left,green] at (1.2,0.75);
			\end{scope}
		\end{tikzpicture}
	}
	\subfloat[DPS TOF \label{subfig:dps_act_tof_hc}]{
		\begin{tikzpicture}
			\begin{scope}[spy using outlines={rectangle,magnification=3,size=8mm,connect spies}]
				\node[inner sep=0pt] {
					\begin{overpic}[width=\tempdima,height=\tempdima]{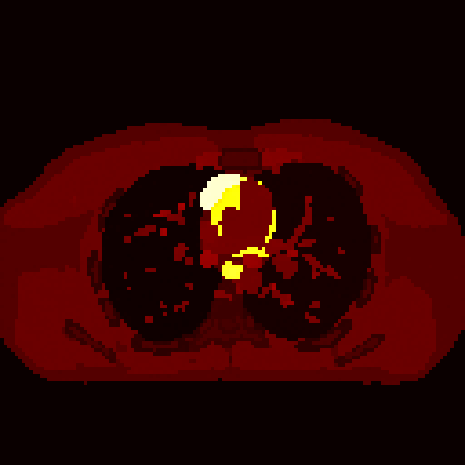}
						\put(1,2){\textcolor{\textcolormetrics}{\scriptsize \acs{PSNR}={31.35}}}
						\put(1,89){\textcolor{\textcolormetrics}{\scriptsize \acs{SSIM}={0.951}}}
					\end{overpic}
				};
				\spy [green] on (0.5,-0.15) in node [left,green] at (1.2,0.75);
			\end{scope}
		\end{tikzpicture}
	}
	\subfloat[DPS2 no TOF \label{subfig:dps2_act_notof_hc}]{
		\begin{tikzpicture}
			\begin{scope}[spy using outlines={rectangle,magnification=3,size=8mm,connect spies}]
				\node[inner sep=0pt] {
					\begin{overpic}[width=\tempdima,height=\tempdima]{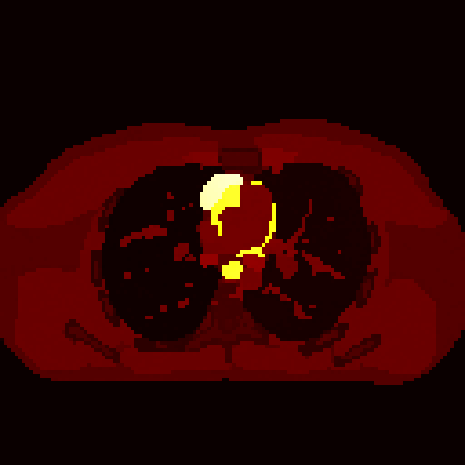}
						\put(1,2){\textcolor{\textcolormetrics}{\scriptsize \acs{PSNR}={30.60}}}
						\put(1,89){\textcolor{\textcolormetrics}{\scriptsize \acs{SSIM}={0.932}}}
					\end{overpic}
				};
				\spy [green] on (0.5,-0.15) in node [left,green] at (1.2,0.75);
			\end{scope}
		\end{tikzpicture}
	}
	\subfloat[DPS2 TOF \label{subfig:dps2_act_tof_hc}]{
		\begin{tikzpicture}
			\begin{scope}[spy using outlines={rectangle,magnification=3,size=8mm,connect spies}]
				\node[inner sep=0pt] {
					\begin{overpic}[width=\tempdima,height=\tempdima]{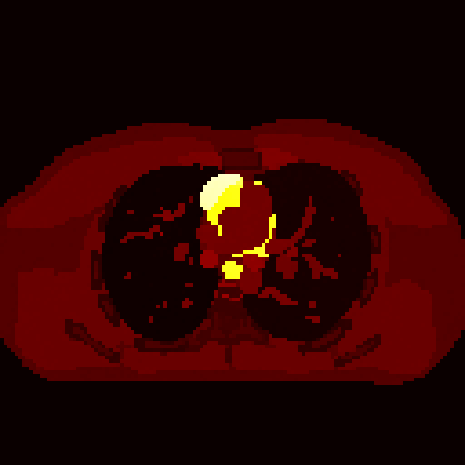}
						\put(1,2){\textcolor{\textcolormetrics}{\scriptsize \acs{PSNR}={30.40}}}
						\put(1,89){\textcolor{\textcolormetrics}{\scriptsize \acs{SSIM}={0.936}}}
					\end{overpic}
				};
				\spy [green] on (0.5,-0.15) in node [left,green] at (1.2,0.75);
			\end{scope}
		\end{tikzpicture}
	}
	
	%%%%% Attenuation images
	\subfloat[\Ac{GT} $\boldmu$ \label{subfig:gt_attn_hc}]{
		\begin{tikzpicture}
			\begin{scope}[spy using outlines={rectangle,magnification=3,size=8mm,connect spies}]
				\node[inner sep=0pt] {
					\includegraphics[width=\tempdima,height=\tempdima]{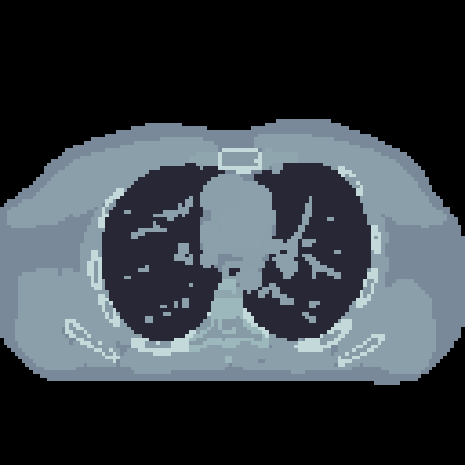}
				};
				\spy [green] on (0.5,-0.15) in node [left,green] at (1.2,0.75);
			\end{scope}
		\end{tikzpicture}
	}
	\subfloat[MLAA no TOF \label{subfig:mlaa_attn_notof_hc}]{
		\begin{tikzpicture}
			\begin{scope}[spy using outlines={rectangle,magnification=3,size=8mm,connect spies}]
				\node[inner sep=0pt] {
					\begin{overpic}[width=\tempdima,height=\tempdima]{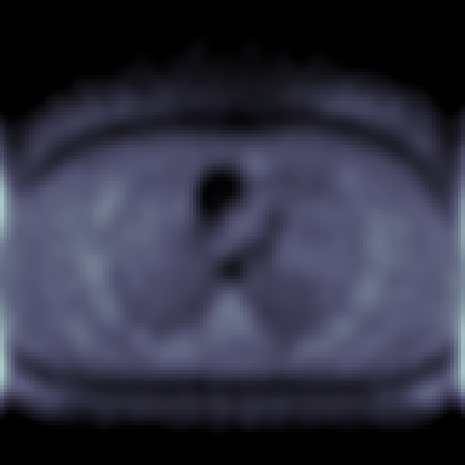}
						\put(1,2){\textcolor{\textcolormetrics}{\scriptsize \acs{PSNR}={11.90}}}
						\put(1,89){\textcolor{\textcolormetrics}{\scriptsize \acs{SSIM}={0.294}}}
					\end{overpic}
				};
				\spy [green] on (0.5,-0.15) in node [left,green] at (1.2,0.75);
			\end{scope}
		\end{tikzpicture}
	}
	\subfloat[MLAA TOF \label{subfig:mlaa_attn_tof_hc}]{
		\begin{tikzpicture}
			\begin{scope}[spy using outlines={rectangle,magnification=3,size=8mm,connect spies}]
				\node[inner sep=0pt] {
					\begin{overpic}[width=\tempdima,height=\tempdima]{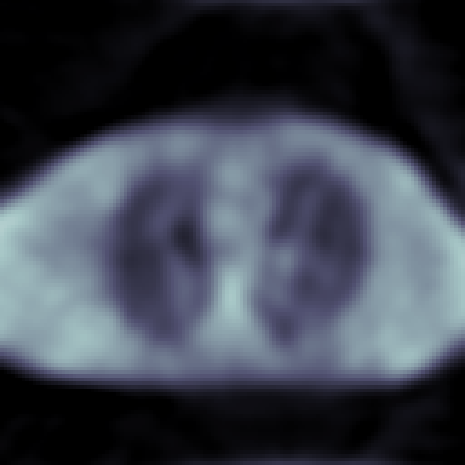}
						\put(1,2){\textcolor{\textcolormetrics}{\scriptsize \acs{PSNR}={16.70}}}
						\put(1,89){\textcolor{\textcolormetrics}{\scriptsize \acs{SSIM}={0.335}}}
					\end{overpic}
				};
				\spy [green] on (0.5,-0.15) in node [left,green] at (1.2,0.75);
			\end{scope}
		\end{tikzpicture}
	}
	\subfloat[DPS no TOF \label{subfig:dps_attn_notof_hc}]{
		\begin{tikzpicture}
			\begin{scope}[spy using outlines={rectangle,magnification=3,size=8mm,connect spies}]
				\node[inner sep=0pt] {
					\begin{overpic}[width=\tempdima,height=\tempdima]{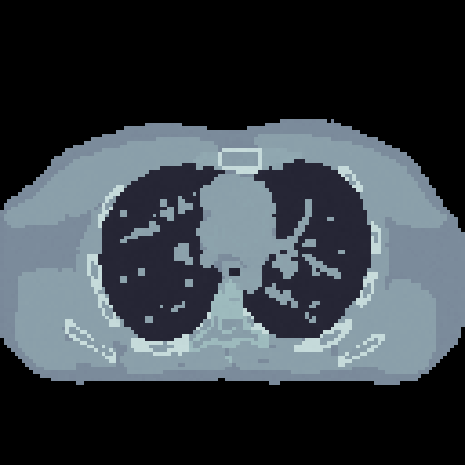}
						\put(1,2){\textcolor{\textcolormetrics}{\scriptsize \acs{PSNR}={23.44}}}
						\put(1,89){\textcolor{\textcolormetrics}{\scriptsize \acs{SSIM}={0.919}}}
					\end{overpic}
				};
				\spy [green] on (0.5,-0.15) in node [left,green] at (1.2,0.75);
			\end{scope}
		\end{tikzpicture}
	}
	\subfloat[DPS TOF \label{subfig:dps_attn_tof_hc}]{
		\begin{tikzpicture}
			\begin{scope}[spy using outlines={rectangle,magnification=3,size=8mm,connect spies}]
				\node[inner sep=0pt] {
					\begin{overpic}[width=\tempdima,height=\tempdima]{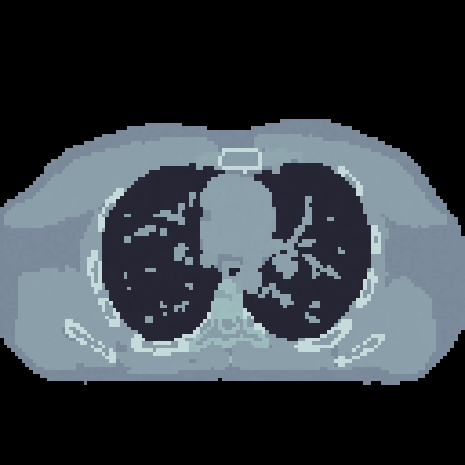}
						\put(1,2){\textcolor{\textcolormetrics}{\scriptsize \acs{PSNR}={25.11}}}
						\put(1,89){\textcolor{\textcolormetrics}{\scriptsize \acs{SSIM}={0.938}}}
					\end{overpic}
				};
				\spy [green] on (0.5,-0.15) in node [left,green] at (1.2,0.75);
			\end{scope}
		\end{tikzpicture}
	}
	\subfloat[DPS2 no TOF \label{subfig:dps2_attn_notof_hc}]{
		\begin{tikzpicture}
			\begin{scope}[spy using outlines={rectangle,magnification=3,size=8mm,connect spies}]
				\node[inner sep=0pt] {
					\begin{overpic}[width=\tempdima,height=\tempdima]{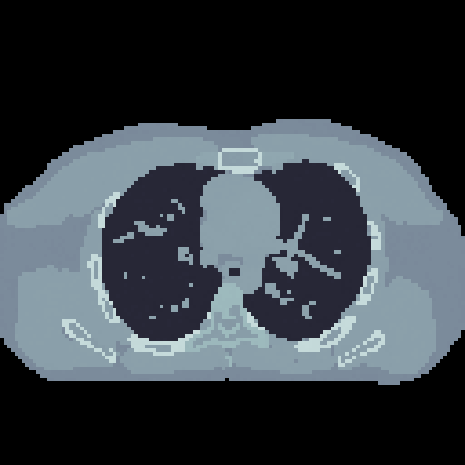}
						\put(1,2){\textcolor{\textcolormetrics}{\scriptsize \acs{PSNR}={23.41}}}
						\put(1,89){\textcolor{\textcolormetrics}{\scriptsize \acs{SSIM}={0.915}}}
					\end{overpic}
				};
				\spy [green] on (0.5,-0.15) in node [left,green] at (1.2,0.75);
			\end{scope}
		\end{tikzpicture}
	}
	\subfloat[DPS2 TOF \label{subfig:dps2_attn_tof_hc}]{
		\begin{tikzpicture}
			\begin{scope}[spy using outlines={rectangle,magnification=3,size=8mm,connect spies}]
				\node[inner sep=0pt] {
					\begin{overpic}[width=\tempdima,height=\tempdima]{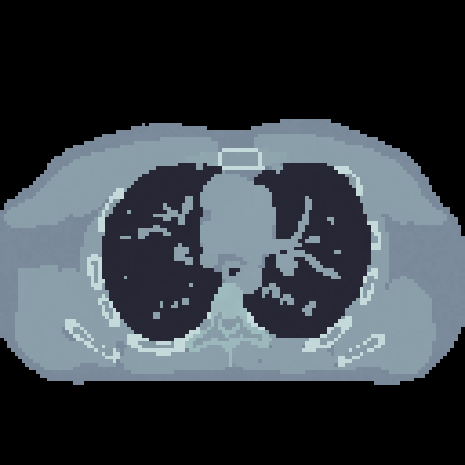}
						\put(1,2){\textcolor{\textcolormetrics}{\scriptsize \acs{PSNR}={24.25}}}
						\put(1,89){\textcolor{\textcolormetrics}{\scriptsize \acs{SSIM}={0.926}}}
					\end{overpic}
				};
				\spy [green] on (0.5,-0.15) in node [left,green] at (1.2,0.75);
			\end{scope}
		\end{tikzpicture}
	}
	
	\caption{
		\Ac{GT} and reconstructed images.
	}\label{fig:recon_hc}
\end{figure*}

\section{Experiments}\label{sec:results}

\subsection{Training, Emission Data Simulation and Evaluation}

All methods (training and reconstruction) were implemented with PyTorch.

We generated a collection of activity-attenuation pairs $\boldx=(\boldlambda,\boldmu)$ using the \ac{XCAT} software \cite{segars20104d} with different morphology and respiratory motion parameters to create a diverse dataset---without tumors included. The images consist of \ac{2D} 128$\times$128 slices with a 3-mm pixel size. The image pairs were used as follows: 4,000 for training, 1,600 for validation and 10 for testing---training and testing slices were extracted from different phantoms. We solved the score-matching problem \eqref{eq:score_matching} using the Adam optimizer with approximately 250 epochs. The training was performed on standardized images, and the standardization was taken into account in the forward model \eqref{eq:poisson}.

%We trained two score functions $\bolds_{\boldtheta}(\cdot,t) \colon \calZ \to \calZ$. The first score uses the proposed model and was trained on matching  $(\boldlambda,\boldmu)$ pairs in order to  account for dependencies between $\boldlambda$ and $\boldmu$, while the second is of the form $\bolds_{\boldtheta}(\boldx, t) = [\bolds_{\boldphi}(\boldlambda, t),\bolds_{\boldpsi}(\boldmu, t)]$, $\boldtheta = (\boldphi,\boldpsi)$, where the two score functions $\bolds_{\boldphi}(\cdot,t),\bolds_{\boldpsi}(\cdot,t) \colon \calX \to \calX$ were trained independently on the activity images and attenuation images respectively---this model assumes that $\boldlambda$ and $\boldmu$ are independent, i.e., $p(\boldx) = p(\boldlambda)\cdot p(\boldmu)$. The  reconstruction using the first model is referred to as \ac{DPS} while reconstruction using the second model will be referred to as \ac{DPS}2.

We trained two score functions $\bolds_{\boldtheta}(\cdot,t) \colon \calZ \to \calZ$. The first score uses the proposed model and was trained on matching  $(\boldlambda,\boldmu)$ pairs in order to  account for dependencies between $\boldlambda$ and $\boldmu$ (referred to as \ac{DPS}), while the second is of the form $\bolds_{\boldtheta}(\boldx, t) = [\bolds_{\boldphi}(\boldlambda, t),\bolds_{\boldpsi}(\boldmu, t)]$, $\boldtheta = (\boldphi,\boldpsi)$, where the two score functions $\bolds_{\boldphi}(\cdot,t),\bolds_{\boldpsi}(\cdot,t) \colon \calX \to \calX$ were trained independently on the activity images and attenuation images respectively---this model assumes that $\boldlambda$ and $\boldmu$ are independent, i.e., $p(\boldx) = p(\boldlambda)\cdot p(\boldmu)$ (referred to as \ac{DPS}2).

The emission raw data $\boldy$ were generated following \eqref{eq:poisson} using pairs $\boldx=(\boldlambda,\boldmu)$ from the testing dataset. We generated \ac{TOF} and non-\ac{TOF} data using an homemade \ac{PET} projector with a 5-mm \ac{FWHM} spatial resolution, a 60-mm \ac{FWHM} temporal resolution and 315 projection angles.   We used $\tau=5\cdot 10^{-1}$ and we ignored background effects ($\boldr = \boldzero_{\calY}$).

The images were reconstructed using \ac{DPS}, \ac{DPS}2 as well as \ac{MLAA}---with 50 outer iterations---for comparison. We used the \ac{PSNR} and \ac{SSIM}---with the \ac{GT} images as references---as figures of merit, computed using the functions  \verb|peak_signal_noise_ratio| and \verb|structural_similarity| from the Python package \verb|skimage.metrics|.

\subsection{Results}

\Ac{GT} and reconstructed images are shown in Figure~\ref{fig:recon_hc}. \Ac{MLAA}-reconstructed images without \ac{TOF} suffer from $\boldlambda$--$\boldmu$ crosstalk as expected, while \ac{MLAA} with \ac{TOF} does a reasonably good job although images suffer from noise amplification and \acp{PVE} due to the intrinsic resolution of the imaging system. On the other hand, \ac{DPS}- and \ac{DPS}2-reconstructed images achieve near perfect-resolution and noise-free reconstruction, even in absence of \ac{TOF} data. In addition, we observe that \ac{DPS}-reconstructed $\boldlambda$ and $\boldmu$ share similar features while some discrepancies can be observed with \ac{DPS}2 (e.g., magnified area in the lungs).

\begin{figure}[!htb]

	\centering
	
	%%%%% Activity
	\subfloat[ Activity $\boldlambda$  \label{subfig:psnr_ssim_act_hc}]{
		\includegraphics[width=0.45\textwidth]{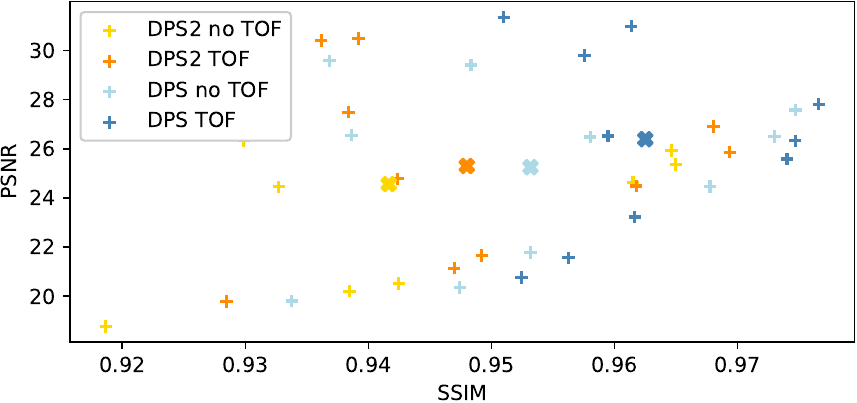}
	}

	%%%%% Attenuation
	\subfloat[ Attenuation $\boldmu$  \label{subfig:psnr_ssim_atn_hc}]{
		\includegraphics[width=0.45\textwidth]{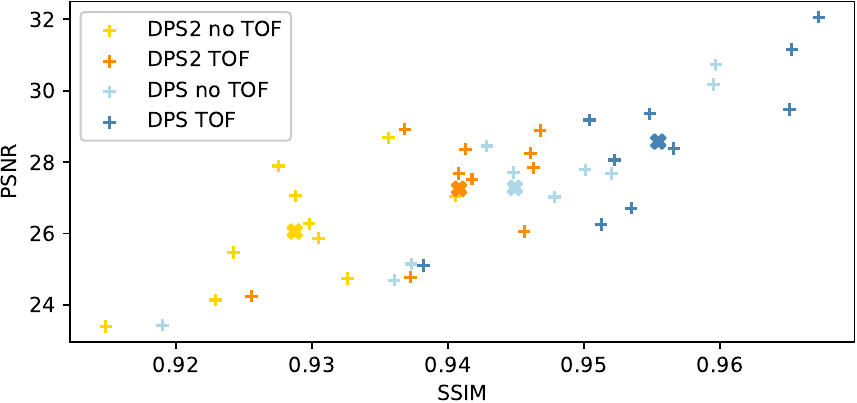}
	}
	
	\caption{
		\Ac{SSIM} vs \ac{PSNR} plots of the \ac{DPS} and \ac{DPS}2 methods for the 10 testing data. The bold crosses represent the mean \ac{SSIM} and \ac{PSNR}  for each reconstruction method.
		}\label{fig:plot_hc}
	
\end{figure}

Figure~\ref{fig:plot_hc} shows \ac{SSIM} vs \ac{PSNR} plots of the \ac{DPS} and \ac{DPS}2 methods for the 10 testing data in the high-count setting---\ac{MLAA} reconstructions were omitted for visibility as they are largely outperformed. We observe that \ac{DPS} outperforms \ac{DPS}2. More particularly,  \ac{DPS} without \ac{TOF} outperforms \ac{DPS}2 with \ac{TOF}, suggesting that the reconstruction benefits of the dependencies between $\boldlambda$ and $\boldmu$.

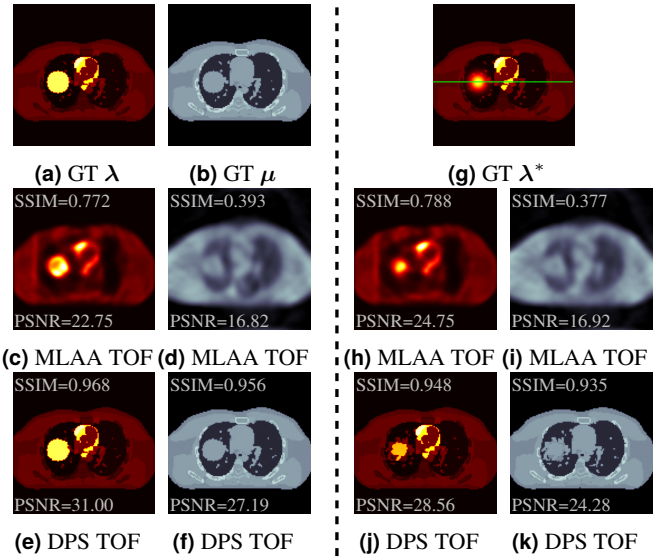
\begin{figure}[htb!]
	\centering
	\begin{tikzpicture}[every node/.style={inner sep=0, outer sep=0}]
		
		% Left block: 6 images in 3 rows and 2 columns
		\node (leftblock) {
			\begin{tabular}{@{}c@{}c@{}}
				\subcaptionbox{\Ac{GT} $\boldlambda$ \label{subfig:gt_act_ood}}{
					\includegraphics[width=\tempdimb,height=\tempdimb]{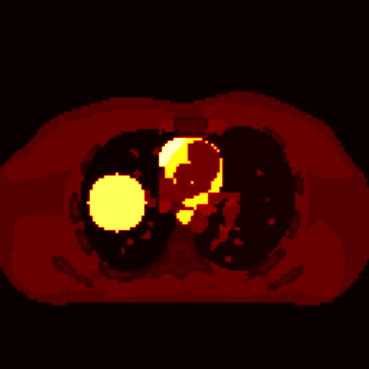}} &
				\subcaptionbox{\Ac{GT} $\boldmu$ \label{subfig:gt_attn_ood}}{
					\includegraphics[width=\tempdimb,height=\tempdimb]{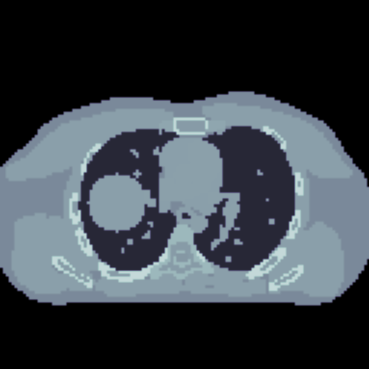}} \\
				\subcaptionbox{\Ac{MLAA} \Ac{TOF} \label{subfig:mlaa_act_tof_ood}}{
					\begin{overpic}[width=\tempdimb,height=\tempdimb]{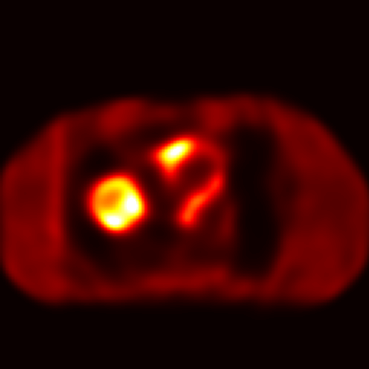}
						\put(-4,2){
							\textcolor{\textcolormetrics}{\scriptsize \acs{PSNR}={22.75} }
						}
						\put(-4,87){
							\textcolor{\textcolormetrics}{\scriptsize \acs{SSIM}={0.772} }
						}
					\end{overpic}
				} &
				\subcaptionbox{\Ac{MLAA} \Ac{TOF} \label{subfig:mlaa_attn_tof_ood}}{
					\begin{overpic}[width=\tempdimb,height=\tempdimb]{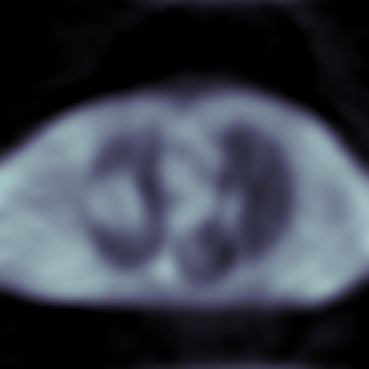}
						\put(-4,2){
							\textcolor{\textcolormetrics}{\scriptsize \acs{PSNR}={16.82} }
						}
						\put(-4,87){
							\textcolor{\textcolormetrics}{\scriptsize \acs{SSIM}={0.393} }
						}
					\end{overpic}
				} \\
				\subcaptionbox{\Ac{DPS} \Ac{TOF} \label{subfig:dps_act_tof_ood}}{
					\begin{overpic}[width=\tempdimb,height=\tempdimb]{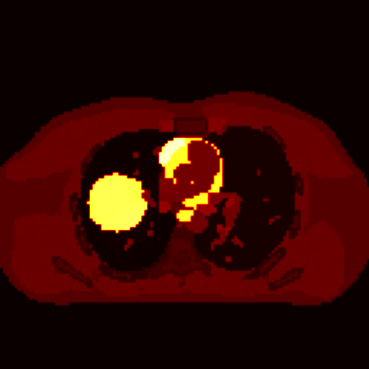}
						\put(-4,2){
							\textcolor{\textcolormetrics}{\scriptsize \acs{PSNR}={31.00} }
						}
						\put(-4,87){
							\textcolor{\textcolormetrics}{\scriptsize \acs{SSIM}={0.968} }
						}
					\end{overpic}
				} &
				\subcaptionbox{\Ac{DPS} \Ac{TOF} \label{subfig:dps_attn_tof_ood}}{
					\begin{overpic}[width=\tempdimb,height=\tempdimb]{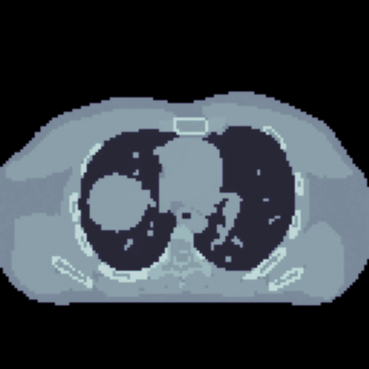}
						\put(-4,2){
							\textcolor{\textcolormetrics}{\scriptsize \acs{PSNR}={27.19} }
						}
						\put(-4,87){
							\textcolor{\textcolormetrics}{\scriptsize \acs{SSIM}={0.956} }
						}
					\end{overpic}
				} \\
			\end{tabular}
		};
		
		% Right block: 5 images (1 row with 1 image, 2 rows with 2 images)
		\node[right=0.4cm of leftblock] (rightblock) {
			\begin{tabular}{@{}c@{}}
				\subcaptionbox{\Ac{GT} $\boldlambda^\ast$ \label{subfig:gt_act_ood_gaussian}}{
					\begin{tikzpicture}
						\node[inner sep=0pt] (o) at (0,0)  { 
							\includegraphics[width=\tempdimb,height=\tempdimb]{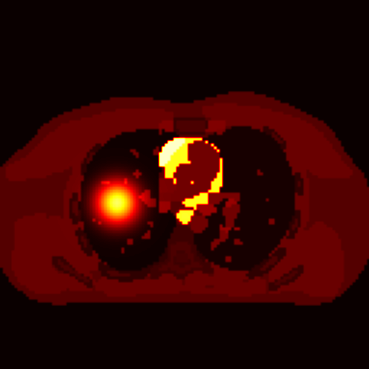}
						} ;
						\draw  [-,>=latex, line width=0.3pt,draw,green]  (0,-0.09) to node [] {} (1.84,-0.09) ;
					\end{tikzpicture}
				} \\
				\begin{tabular}{@{}c@{}c@{}}
					\subcaptionbox{\Ac{MLAA} \Ac{TOF} \label{subfig:mlaa_act_tof_ood_gaussian}}{
						\begin{overpic}[width=\tempdimb,height=\tempdimb]{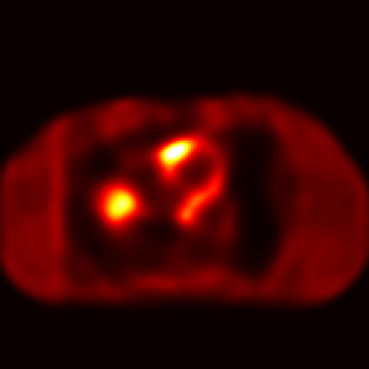}
							\put(-4,2){
								\textcolor{\textcolormetrics}{\scriptsize \acs{PSNR}={24.75} }
							}
							\put(-4,87){
								\textcolor{\textcolormetrics}{\scriptsize \acs{SSIM}={0.788} }
							}
						\end{overpic}	
					} &
					\subcaptionbox{\Ac{MLAA} \Ac{TOF} \label{subfig:mlaa_attn_tof_ood_gaussian}}{
						\begin{overpic}[width=\tempdimb,height=\tempdimb]{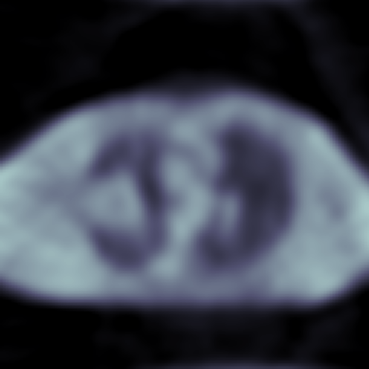}
							\put(-4,2){
								\textcolor{\textcolormetrics}{\scriptsize \acs{PSNR}={16.92} }
							}
							\put(-4,87){
								\textcolor{\textcolormetrics}{\scriptsize \acs{SSIM}={0.377} }
							}
						\end{overpic}
					} \\
				\end{tabular} \\
				\begin{tabular}{@{}c@{}c@{}}
					\subcaptionbox{\Ac{DPS} \Ac{TOF} \label{subfig:dps_act_tof_ood_gaussian}}{
						\begin{overpic}[width=\tempdimb,height=\tempdimb]{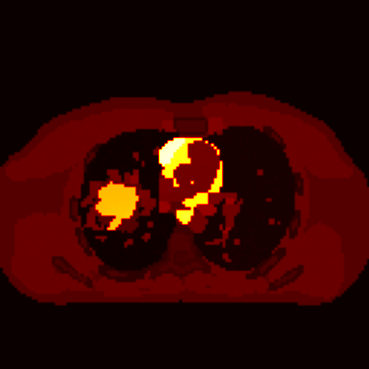}
							\put(-4,2){
								\textcolor{\textcolormetrics}{\scriptsize \acs{PSNR}={28.56} }
							}
							\put(-4,87){
								\textcolor{\textcolormetrics}{\scriptsize \acs{SSIM}={0.948} }
							}
						\end{overpic}
					} &
					\subcaptionbox{\Ac{DPS} \Ac{TOF} \label{subfig:dps_attn_tof_ood_gaussian}}{
						\begin{overpic}[width=\tempdimb,height=\tempdimb]{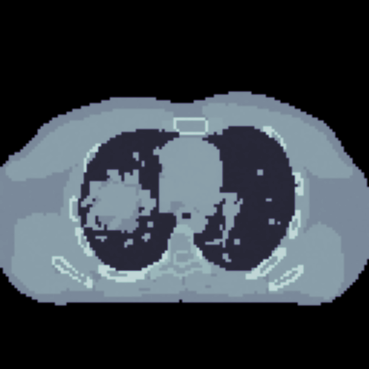}
							\put(-4,2){
								\textcolor{\textcolormetrics}{\scriptsize \acs{PSNR}={24.28} }
							}
							\put(-4,87){
								\textcolor{\textcolormetrics}{\scriptsize \acs{SSIM}={0.935} }
							}
						\end{overpic}
					} \\
				\end{tabular} \\
			\end{tabular}
		};
		
		% Dashed line separating left and right blocks
		\draw[dashed, line width=1.25pt] ($(leftblock.east) + (0.3cm, -3.65cm)$) -- ($(leftblock.east) + (0.3cm, 3.65cm)$);
		
	\end{tikzpicture}
	\caption{
		\Acs{OOD} reconstructions using \ac{MLAA} and  \ac{DPS}, both with \ac{TOF}, where the tumor in the \ac{GT} activity images takes the form of a uniform disk and of a Gaussian function---the \ac{GT} attenuation map $\boldmu$ is identical in both scenarios.
		}
	\label{fig:ood}
\end{figure}

\begin{figure}[!htb]
	
	\centering
	
	\includegraphics[width=0.45\textwidth]{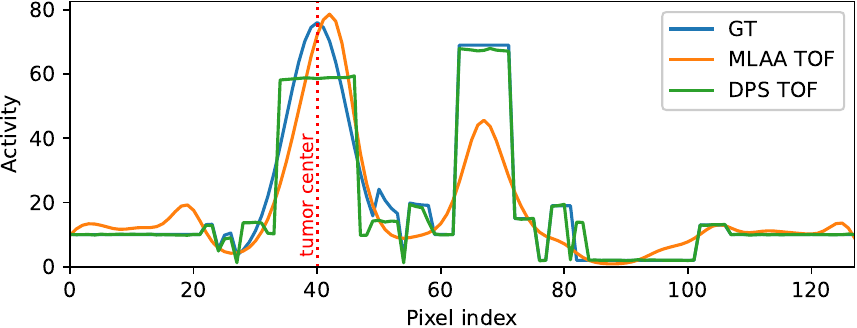}
	
	\caption{
		Activity profiles of \ac{OOD} activity images. The profiles were drawn along the green line in Figure~\ref{subfig:gt_act_ood_gaussian}.
		}\label{fig:ood_profiles} 
	
\end{figure}

The near perfect results achieved by \ac{DPS} and \ac{DPS}2 may raise questions about potential over-fitting due to potential lack of diversity in the \ac{XCAT}-generated training data. We therefore considered two \ac{OOD} cases with a tumor in the lungs, illustrated in Figure~\ref{subfig:gt_act_ood}, Figure~\ref{subfig:gt_attn_ood} and Figure~\ref{subfig:gt_act_ood_gaussian}. The tumor takes the form of a uniform disk the first activity image (denoted $\boldlambda$) and in the attenuation map (denoted $\boldmu$), while it is modeled with a \ac{2D} Gaussian kernel in the second activity image (denoted $\boldlambda^\ast$). We then performed applied \ac{MLAA} and \ac{DPS} to the corresponding \ac{TOF} data. In the first case, \ac{DPS} is able to reconstruct the tumor in the activity and the attenuation 
%when it is a uniform disk in the \ac{GT} activity image 
(cf. Figure~\ref{subfig:dps_act_tof_ood} and Figure~\ref{subfig:dps_attn_tof_ood}). In the second case, the \ac{DPS} reconstruction is somehow capable of producing a ``piecewise constant'' tumor in the activity that mimics the Gaussian shape of the \ac{GT} (cf. Figure~\ref{subfig:dps_act_tof_ood_gaussian} and profiles  in Figure~\ref{fig:ood_profiles}). This behavior results of the training on \ac{XCAT} images which are piecewise constant. In addition, since the score was trained on matching image pairs, the reconstructed tumor in the attenuation map inherits the same artifacts.

\section{Discussion}\label{sec:discussion}

The findings highlight the efficacy of \ac{DPS} for \ac{JRAA} in \ac{PET} using emission data only, particularly in non-TOF settings where conventional methods often falter due to activity–attenuation crosstalk. Unlike \ac{MLAA}, the \ac{DPS} framework integrates prior knowledge through the learned joint \ac{PDF} of activity and attenuation, ensuring consistency between the reconstructed images and effectively addressing crosstalk challenges. The experimental results confirm that \ac{DPS} can generate high-quality, noise-free reconstructions, even under non-\ac{TOF} conditions, and that it surpasses its independently trained variant, DPS2, by leveraging activity--attenuation dependencies. Similar observations were made in multi-energy \ac{CT} reconstruction \cite{vazia2024diffusion}.

Nevertheless, this study acknowledges certain limitations. The models were exclusively trained on \ac{2D} \ac{XCAT} phantom data, which, despite their widespread use in simulation studies, lack the anatomical diversity and complexity of real-world patient datasets. This constraint potentially limits the generalizability of the model to clinical scenarios. Furthermore, the extension of this framework to \ac{3D} volumes remains a critical next step. We are currently working on extending this method by incorporating real patient data and using data compression techniques to handle \ac{3D} data with encouraging results. These advancements are essential to validate the clinical applicability of \ac{DPS} and to realize its potential in routine \ac{PET} imaging workflows.

\section{Conclusion}\label{sec:conclusion}

This study demonstrates the promise of \ac{DPS} as an innovative tool for \ac{JRAA} in \ac{PET} imaging. By directly modeling the dependencies between activity and attenuation, \ac{DPS} achieves superior performance, particularly in non-\ac{TOF} scenarios. While the results represent a significant advancement over traditional \ac{MLAA} methods, ongoing work aims to extend the approach to \ac{3D} data and real-world clinical conditions with encouraging preliminary findings.

\section*{Acknowledgment}
This work was supported by the French National Institute of Health and Medical Research (Inserm), the French Institute for Research in Computer Science and Automation (Inria), the French National Research Agency (ANR) under grant No ANR-20-CE45-0020 and by France Life Imaging under grant No ANR-11-INBS-0006.

%-------------------------------------------------------------------------------------------

\setlength\bibitemsep{1pt}
\AtNextBibliography{\footnotesize}
{	\setstretch{0.1}							% decreases the space between lines
	
	\printbibliography							% add the references
}

\end{document}